# A Perspective on Blockchain Smart Contracts:

Reducing Uncertainty and Complexity in Value Exchange

Henry Kim and Marek Laskowski
Blockchain Lab @ York
Schulich School of Business, York University
Toronto, Canada

*Abstract*— The blockchain constitutes a technology-based, rather than social or regulation based, means to lower uncertainty about one another in order to exchange value. However, its use may very well also lead to increased complexity resulting from having to subsume work that displaced intermediary institutions had performed. We present our perspective that smart contracts may be used to mitigate this increased complexity. We further posit that smart contracts can be delineated according to complexity: Smart contracts that can be verified objectively without much uncertainty belong in an inter-organizational context; those that cannot be objectively verified belong in an intra-organizational context. We state that smart contracts that implement a formal (e.g. mathematical or simulation) model are especially beneficial for both contexts: They can be used to express and enforce inter-organizational agreements, and their basis in a common formalism may ensure effective evaluation and comparison between different intra-organizational contracts. Finally, we present a case study of our perspective by describing Intellichain, which implements formal, agent-based simulation model as a smart contract to provide epidemiological decision support.

## I. Introduction

Many in the business world are hailing the blockchain as the next big technological innovation—placing it nearly at par with the Internet. Marc Andreesen, who as a co-writer of Mosaic and a co-founder of Netscape is truly a pioneer of the Web, said of blockchain: "when we're sitting here in 20 years, we'll be talking about [bitcoin and blockchain technology] the way we talk about the Internet today" [1]. The Tapscotts argue that inasmuch as double-entry bookkeeping enabled the rise of capitalism, this distributed "new digital ledger of economic transactions can be programmed to record virtually everything of value and importance to humankind: birth and death certificates, marriage licenses, deeds and titles of ownership, educational degrees, financial accounts, medical procedures, insurance claims, votes, provenance of food, and anything else that can be expressed in code" [2].

Within the Tapscotts' effusive description is the concept that fundamentally a blockchain is used to implement a ledger—*merely* a ledger. At their onsets, the Internet was not merely a communications network, and the Web was not merely a document database. We state this not to minimize the possible impact of the blockchain, but rather to make the case that unlike the Internet and the Web, blockchain is conceptually simple enough to frame within a recognizable context. Warburg saliently provides such context: "There is a new, technological institution that will fundamentally change how we exchange value, and it's called the blockchain. Now, that's a pretty bold statement… remember that while blockchain technology is relatively new, it's also a continuation of a very human story, and the story is this. As humans, we find ways to lower uncertainty about one another so that we can exchange value" [3].

It is this context of technology to reduce uncertainty—the context within which the blockchain can be placed—that we explore further in this paper.

## II. Uncertainty and Complexity

When we think philosophically about the blockchain and reducing uncertainty, we encounter the inherent trade-off that entails increased complexity. This follows from the Theory of Bounded Rationality by Nobel Laureate Herbert Simon [4] which states that the human mind's processing capacity is small relative to the size of the problems requiring processing for an objective solution. Bounded rationality compels humans or processors to seek techniques to reduce information, task, and coordination complexity to solve problems. As humans and processors find it effective to act collectively to reduce these complexities, organizational structures form.

However, as these structures form, they simultaneously and inevitably face increased uncertainty [5]. The trade-off for abstracting complexities to another organization—not having to know what information is required by them, how they perform tasks, and how they coordinate—and contracting to them as a form of value exchange is that your organization is uncertain of their own information, tasks, and coordination. For example, if your organization used to be responsible for manufacturing and customer service but you decide to outsource the servicing, the price for not having to deal with the complexities of customer service anymore is that you are more uncertain of how the outsourcer actually services your customers. The inverse then should also hold: As humans and processors break down organizational structures to reduce uncertainty, increased complexity would result. If your company decides to integrate customer service back to the organization, then the price of being more certain about how your customers are serviced is that you have to bear the complexities of providing customer service.

Let's consider a simple digital goods example. Say that your organization currently sells subscriptions for an online stock tips newsletter. You are unhappy with the flat fee that payment processors like Visa and Paypal take because you believe that it is too high a percentage of the monthly subscription fee. You decide to modify your current Website and exclusively only take orders in bitcoins. You are happy that the bitcoin network ensures that buyer payments are received without fraud and at higher margins; the network gives you even greater certainty of successful payment processing. Changing your business also forces you to become more certain about how you take orders. You are surprised however when you realize how much these intermediary institutions allowed you to abstract away the complex details of payment processing you now have to figure out—e.g. monthly billing and assuming credit risk.

Bottom line: Blockchain is a technological innovation that reduces uncertainty of value exchange but it may very well also lead to increased complexity resulting from having to subsume work that displaced intermediary institutions had performed.

### III. COMPLEXITY AND SMART CONTRACTS

Let's delve further into complexity using the bounded rationality model. According to Simon [4], an organization structures to decrease complexity in the following manner: an organization structures to accommodate *division of labor* to decrease information and task complexity; then, to product or functional orientation within the organization by structuring of *near decomposable organizational units*, which reduce task coordination complexity by enabling highly coupled coordinated tasks within the unit and lowly coupled tasks between units; and finally, to *contracting* to extra-organizational near decomposable units—i.e. another organization. In contracting, "information complexity is reduced to a price; and task complexity, to contractual terms with a near decomposable unit, the contractor" [6].

If the contracting organization had been willing to contract a set of tasks to the contractor, those tasks must have been performed especially productively by the contractor. Presumably then, the evolution from division of labor to intra-organizational near decomposable units with respect to those set of tasks must have occurred, and with particular effectiveness, within the contractor. So, if an organization determined to use the blockchain eliminates an intermediary contractor, then it must find alternative means to provide division of labor and near decomposable units' capabilities nearly as effectively as the displaced contractor. The elegance of blockchain design is that one of its critical features, *smart contracts*, represents means to provide such division of labor and near decomposability; that is, smart contracts offer technological means to mitigate complexity in blockchain use.

In the Industrial Age, division of labor delineated responsibilities between workers and machines. Smart contracts, as "applications [that execute on the Ethereum blockchain] that run exactly as programmed without any possibility of downtime, censorship, fraud or third party interference" [7] constitute the machine aspect of the blockchain-age division of labor. In fact, the promise of smart contracts is that the division of labor between workers and computer applications as practiced by a displaced contractor would be replaced exclusively by a workerless, reliable, and censor, fraud, and tamper proof "labor force" of smart contracts. Furthermore, smart contracts can be carefully engineered to exhibit characteristics of near decomposability; that is, smart contracts can be grouped into a near decomposable unit within which there is great task coordination, and less between. Ultimately, a system of smart contracts may function as the name implies: "pieces of software that represent a business arrangement and execute themselves automatically under pre-determined circumstances" [8].

There are the two possible configurations for a system of smart contracts:

- A smart contracts system serves as a near decomposable unit within an organization. Its main responsibility would be to perform, govern, and coordinate those blockchain-related tasks that are primarily internal to the organization. It would provide interface capabilities (like an API) so that inter-organizational tasks can be performed. An individual smart contract in this system may look very much like a traditional computer program in that it would be expressed procedurally. We posit this because as a unit within the overall organization this system ought to have visibility and access to various non-blockchain aspects of the organization's data, business processes, systems, and workers. As a result, the contracts can orchestrate the details of how to perform tasks, and this would best be done through procedural programming.

- A smart contract system may itself acts as a contractor. We envision several forms. One, it may be operated by the two parties of a bilateral contract to govern and enforce that contract. Two, it may be operated by a consortium comprising of parties in a multilateral contract. Or, three, it may be operated by a third-party who offers superior transparency, impartiality, lack of power-brokering, and efficiency than a traditional intermediary displaced via blockchain use. Regardless of the form, this contractor ought not to have intimate access to contracting organizations' systems, especially their systems off of the blockchain. Instead, it should perform and coordinate tasks via these organizations' interfaces (as mentioned above). These smart contracts should declare what tasks need to be performed by contracting organizations without necessarily specifying how. That is, these contracts should be programmed more declaratively, specifying contractual obligations and how they are to be enforced without detailing how contracting organizations are to meet these obligations.

Bottom line: Smart contracts may be used to mitigate the increased complexity resulting from eliminating intermediaries through blockchain use. Bounded rationality states that organizations structure towards division of labor, near decomposition, and contracting to decrease complexity. Smart contracts represent programmatic means to efficiently apply these structures for value exchange.

## IV. UNCERTAINTY AND SMART CONTRACTS

Let's go back to the example of your newsletter company. Recall that you faced increased complexity for monthly billing and assuming credit risk. You realize that you either have to ask your customers every month to pay or rely upon them to remember to pay every month. Before, the payment processors took care of these details. More importantly, you realize that you have now assumed credit risk that the payment processors used to be responsible for. If you cancel a subscription because a customer has not paid, you run the risk that you are losing a loyal customer who happened to have insufficient funds on the due date or simply forgot. If you do not cancel and give more time to receive payment, you may be serving a free-loader who has no intention to pay. Before, the payment processor stepped in to assume credit risk. You did not have to cancel until the customers told you to because as long as they didn't cancel the payment processors paid you. The matter of extracting payment then was between the processor and the customer, not you. Your customers now also assume a risk: what if you exploit automated unsupervised monthly billing by double-billing or billing for extra months? You realize that your old payment processor had provided risk protection for not only you but also for your customers.

Smart contracts give you means to deal with these risks without the payment processor. You can specify a set of smart contracts for your risk protection such as monthly price, subscription renewal period, penalty for late payment, and terms for dropping the customer. Other matters that can allay risk for the customer such as caps on payment, frequency of newsletters, as well as terms for dropping the newsletter service can also be specified as smart contracts. All these matters can be verified objectively, except for the ones specifying dropping customers or the service, which require subjective verification. It can be objectively specified and verified that renewal of the subscription occurs once a year and that a newsletter must be published every trading day morning at 7AM, but it is more subjective to specify and verify the standards on the quality of the newsletters to justify dropping the service or changing the terms of the service like monthly price. This dichotomy between objective and subjective smart contracts can be framed within a context of smart contracts and uncertainty.

We have stated that smart contracts can be included within extra-organizational system of contracts or intra-organizational near decomposable system of contracts. The newsletter example illustrates that an importation factor in determining where a smart contract is included is the certainty with which it can be verified. Objective contracts should be included with the contractor; subjective contracts should be within the purview of the organization.

So, maybe you specify a smart contract to verify that your newsletter is of high quality, and hence you believe you are warranted in asking for more for monthly subscriptions because you can demonstrate that the stock tips lead on average to greater investment income for your customers. In contrast, a particular customer may wish to cancel their subscription without penalty even though it is not yet the yearly renewal period because they believe the newsletter is of low quality, having experienced investment losses acting upon your tips. You and that customer may develop smart contracts specifying the quality of your newsletters that will yield contradicting conclusions.

Bottom line: Smart contracts can be delineated according to uncertainty. Smart contracts that can be verified objectively without uncertainty belong in an extra-organizational system of smart contracts. This establishes *agreements* between two or more parties. Smart contracts that can only be verified subjectively with uncertainty belong in an intra-organizational system of smart contracts. This establishes *models for one party*.

## V. SMART CONTRACTS AS FORMAL MODELS

Figure 1 compares agreement versus one-party smart contracts.

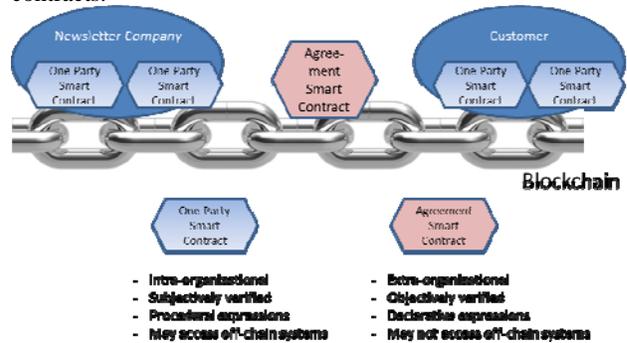

Fig. 1. Agreement vs. one-party Smart Contracts

Let us now adopt some assumptions. Fundamentally, a system of smart contracts represents a model of a phenomenon represented using blockchain constructs. We further constrain these models to be formal. That is, a well-formed formal model yields a unique set of interpretations or inferences. Or minimally, a formal model yields a bounded set of interpretations or inferences. The key difference then between agreement and one-party smart contracts as models is that agreement contracts require multiple parties to "agree upon" a formal model, whereas one-party contracts can be developed by one party. In effect, an agreement contract models a phenomenon identically for all relevant parties, whereas two parties for whom a phenomenon is relevant may develop two one-party contracts of the same phenomenon in very different ways. Referring back to the newsletter service example, you and your customers may develop a multi-party agreement smart contract of terms of the service which can be interpreted (i.e. inferred, verified, etc.) unambiguously by all parties. At the same time, you may develop a one-party smart contract of your service quality where the interpretation of the contract is quite different and likely more flattering from the interpretation of your customer's one-party smart contract of your service quality.

An interesting question is whether the two one-party contracts of service quality ought to be visible to both parties. So, should the newsletter service see their customer's less than flattering smart contract model of their own service quality? Should the customer see the service's glowing smart contract model of service quality? To answer this, recall Warburg's provocative point: "As humans, we find ways to lower uncertainty about one another so that we can exchange value" [3].

Bottom line: The ability to evaluate different parties' smart contract models of the same phenomenon increases transparency and lowers uncertainty of the value exchange. Moreover, it is desirable to make the model transparent yet hide some of the data required to interpret the model. Or in the parlance of the blockchain, the smart contract can be publicly viewable, but the data required for verification may or may not be open to all.

In the next section, we discuss some works that develop formal models to underlie their blockchain smart contracts, and present a brief case study of the *Intellichain* formal modeling endeavor.

## VI. CASE STUDY

Arguably, any endeavor that designs smart contracts assumes a model of a domain. However, in the current literature, most of these models are not formal. We define a formal model as a model expressed in a formal language such that a valid expression in that language entails an unambiguous interpretation and automatable inference. A valid smart contract expression encoded in say, Solidity, can be uniquely interpreted by a compiler or interpreter. However, the conceptual data and process models of that domain for which an entailment is implemented in Solidity is, more often than not, informal. As an example, an informal model could be a set of business rules proprietary to a given company; a formal model would be a mathematical formula for calculating amounts exchanged in an interest rate swap.

Exactly such a formula is implemented in the first proof-of-concept for R3's Corda [9]. In this implementation using the Corda blockchain, interest rate swap agreements between different banking partners are registered on the blockchain and smart contracts automatically execute terms of these agreements. However, specific terms of these agreements unnecessary for parties not involved in a swap to view are not made transparent to these parties.

In the ModelChain framework [10], formal statistical predictive models are implemented as smart contracts. Specifically, each node in a permissioned healthcare blockchain would execute smart contracts to calculate and publish local statistical model estimation parameters to be used to calculate parameters for a global predictive model without making local patient data transparent. Furthermore, smart contracts could then execute the global model on the local data to make predictions about "local" patients, only making broadly visible worthwhile results or insights that still maintain the privacy of these patients.

Similar to ModelChain in its applicability for public health is Intellichain, which can be used to assess public health intervention strategies through the use of a decision support framework with the core simulation component implemented as a smart contract.

Epidemiological data inputs into the system can come from World Wide Web as well as mobile devices including smartphones and Internet of Things (IoT) devices. In the context of this proof-of-concept, all this data is anonymized, aggregated, and stored using a relatively simple mechanism within a smart contract, implemented using the Solidity programming language. This smart contract also implements the formal, well-known SIR (Susceptible, Infected, Recovered) agent-based simulation model [11] used in epidemiology to describe the spread of infectious disease. The SIR model can also be used to describe the spread of information in a network of individuals [12]. Analogous to ModelChain, estimates for model parameters are calculated via aggregating the collected data in order to execute the SIR model.

The SIR model as with many Agent Based Models [13] features discrete models for locations, agents, disease compartments, and time [14][15]. Probability of disease transmission from an infected to susceptible agent is modelled as a parameter $\beta$, which is used in Bernoulli trials to determine if the disease spreads from an infected individual to a susceptible individual in close contact. The Infectious period as well as $\beta$ are drawn from uniform distribution, and their values are based on published estimates in literature [14][16]. The movement behavior of individuals is modeled as a Markov process based on imputed observations of transitions between locations. Each individual is parameterized with a number of contacts per time step, also based on observations or available data for that individual. For further implementation details please refer to https://github.com/professormarek/Intellichain and for deployment details and system diagram refer to [17].

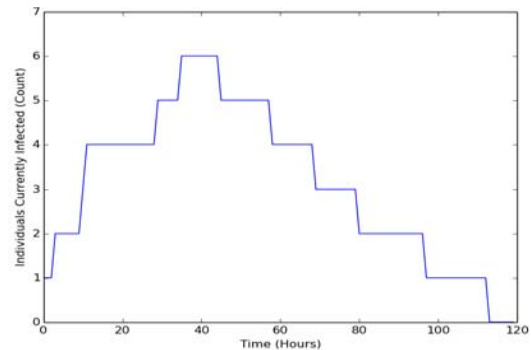

Fig. 2. Number of currently infected individuals on each hour of a simulation.

The simulation results can be interpreted using standard analytical tools such as graphs and charts. One such graph is Figure 2 which shows the number of currently infected individuals as a function of time during a simulation with particular initial conditions carried out on the test network.

In addition to a traditional web-interface presenting aggregate graphs, charts, and statistics, the framework includes a Virtual Reality "Decision Support Table" implemented using the Google Daydream VR kit[1] and Unity3D[2]. VR represents an intuitive and explicit means of visualizing data and simulation outcomes during the decision phase, and can be used for knowledge translation to convey the nature of simulation based modelling to stakeholders and the public during knowledge dissemination in the action phase.

A demonstration of how VR can support visualizing and communicating simulation and decision support results has been published to the Google Play store for Daydream compatible devices. This app permits users to experience the simulation described by Figure 2, as carried out earlier on the test network. Both Figure 2 and the VR app communicate the important epidemiological principle of a super-spreader infecting several secondary cases early on in the outbreak. These cases in turn infect others later on. Once the app is launched, the user can look around in 3D space as well as navigate using the Daydream VR kit's provided controller to view the simulation from a multitude of perspectives and speeds. The virtual decision table is visible in Figure 3 which is a screen capture of the VR app as a simulation is playing out.

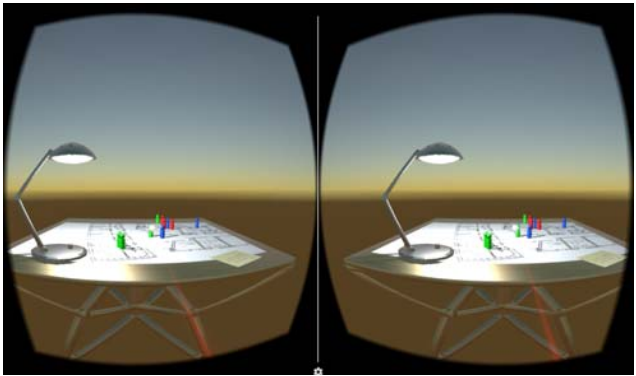

Fig. 3. Screen capture of decision table VR simulation demonstrating the images shown to each eye.

Although at first glance this work appears to be an instance of using models for one party, in practice, there are potentially multiple stakeholders, each with their own domain specific model implementations. This suggests the potential for an Agreement Smart Contract in order to be able to compare results and outcomes between independently developed domain specific models. Figure 4 depicts one such scenario, in which an Agreement contract C is being used to compare the assessment made by two different Simulation contracts (A & B). The agreement contract needs to be able to inspect and reason whether the evidence provided by contract A is compatible with the evidence provided by contract B based on their respective formal models. This inspection is possible, though not guaranteed, if the modelling formalisms used for A is the same as in B.

---

[1] https://vr.google.com/daydream/
[2] https://unity3d.com/

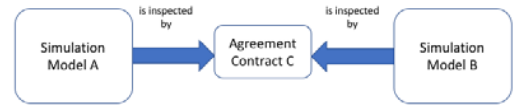

Fig. 4. In a multi-stakeholder scenario an Agreement Contract, C, would have to inspect the underlying formal models of contracts A and B in order to determine whether the evidence provided by A and B is consistent.

## VII. CONCLUSIONS

We have stated the following principles about the potential evolution of smart contracts vis-à-vis the blockchain.

- The blockchain is a technological innovation that reduces uncertainty of value exchange but it may very well also lead to increased complexity resulting from having to subsume work that displaced intermediary institutions had performed.

- Smart contracts may be used to mitigate the increased complexity resulting from eliminating intermediaries through blockchain use. Bounded rationality states that organizations structure towards division of labor, near decomposition, and contracting to decrease complexity. Smart contracts represent programmatic means to efficiently apply these structures for value exchange.

- The ability to evaluate different parties' smart contract models of the same phenomenon increases transparency and lowers uncertainty of the value exchange. Moreover, it is desirable to make the model transparent yet hide some of the data required to interpret the model. Or in the parlance of the blockchain, the smart contract can be publicly viewable, but the data required for verification may or may not be open to all.

To that last point, this ability to evaluate, commonly interpret, and maintain transparency of different parties' smart contracts is enhanced if the contracts implement formal models—whether they are mathematical, logical, or simulation-based. As a case study of this assertion, we detail Intellichain, which uses smart contracts to implement a formal, agent-based simulation model on the blockchain. As long as different parties adhere to a common modelling formalism, conclusions drawn from one party's simulation can be evaluated "apples to apples" with those from another party.